\documentclass[%
 amsmath,amssymb,
 aps,
]{revtex4-2}

\usepackage{graphicx}
\usepackage{dcolumn}
\usepackage{bm}

\usepackage{txfonts}
\usepackage{comment}

\graphicspath{{./figures/}}
\usepackage{color}
\definecolor{violet}{rgb}{0.56, 0.0, 1.0}

\begin{document}

\preprint{APS/123-QED}

\title{Synchronization approach to achieving maximum power and thermal efficiency for weakly-coupled low-temperature-differential Stirling engines}

\author{Songhao Yin}
\email{4894715006@edu.k.u-tokyo.ac.jp}
\author{Hiroshi Kori}
\author{Yuki Izumida}
\affiliation{
Graduate School of Frontier Sciences, The University of Tokyo, Kashiwa 277-8561, Japan
}

\date{\today}

\begin{abstract}
Low-temperature-differential (LTD) Stirling engines are heat engines that can operate autonomously with a slight temperature difference between low-temperature heat reservoirs and are thus expected to contribute to a sustainable society. A minimal dynamical-system model with only two variables has been proposed to explain the principle of autonomous rotational motion caused by temperature differences, and the maximum efficiency of the engine was formulated [Y. Izumida, Europhys. Lett. {\bf 121}, 50004 (2018); Phys. Rev. E {\bf 102}, 012142 (2020)]. This paper aims to clarify the coupling effects on the dynamics, power, and thermal efficiency of a pair of weakly coupled LTD Stirling engines and formulate the maximum thermal efficiency of the coupled system in the quasilinear response regime. We show that the dependence relation between the effective frequency difference and the coupling strength is characterized by a hysteresis, which comes from different kinds of bifurcations in the process of increasing and decreasing the value of the coupling strength. Then, by generalizing thermodynamic fluxes and forces and their quasilinear relations for engines under weak coupling, we show that the coupling improves the power exerted against the load torques and the thermal efficiency. We further show that their maximum values are achieved when the engines are synchronized. Since the thermal efficiency depends on the frequency difference, the dependence of thermal efficiency on the coupling strength is also characterized by a hysteresis. Finally, the load torque that achieves the maximum thermal efficiency of the coupled system is formulated.
\end{abstract}

\maketitle


\section{Introduction}
A heat engine is a system that uses thermal energy from a high-temperature heat reservoir to extract positive work. According to the second law of thermodynamics, a low-temperature heat source is required to discard part of the thermal energy to extract positive work from a heat engine. Low-temperature differential (LTD) Stirling engines, which can rotate autonomously with only a slight temperature difference between low-temperature heat reservoirs, are attracting significant attention as an elemental technology to realize a sustainable society \cite{Senft2010,Senft2000,kongtragool2003review}. From this perspective, it is vital to understand the dynamical characteristics of LTD Stirling engines through appropriate mathematical modeling \cite{robson2007modelling,craun2018control}. A nonlinear dynamics model has been proposed to explain the loss of rotational motion of LTD Stirling engines, which was found to be caused by a homoclinic bifurcation \cite{izumida2018nonlinear}.

Another important issue for the LTD Stirling engines is thermal efficiency. In \cite{izumida2020quasilinear}, one of the authors demonstrated that the engine's rotational state is in a quasilinear response regime where the thermodynamic fluxes show a linear dependence on the thermodynamic forces and formulated the maximum efficiency of the engine based on the fact that the response coefficients of the quasilinear relations are symmetric, which is similar to Onsager symmetry in linear irreversible thermodynamics. However, the power extracted from a single LTD Stirling engine is quite limited, thus it is desirable to operate a population of Stirling engines to extract adequate work for practical purposes. Methods that achieve maximum efficiency by properly controlling a population of Stirling engines then turn out to be important.

Synchronization is a self-organized phenomenon in which oscillators align their rhythms through interaction and is widely observed in natural and artificial systems \cite{Pikovsky,kuramoto1984chemical}. A natural question that would be raised is whether synchronization through coupling between the LTD Stirling engines can improve the total power and thermal efficiency. If it does, then higher power and thermal efficiency can be achieved simply by allowing the engines to interact with each other. Although experimental studies on the synchronization of LTD Stirling engines have been conducted \cite{kada2014synchronization,migimatsu2017experimental}, the effects of synchronization on power and thermal efficiency have not yet been clarified theoretically.

This paper aims to clarify the coupling effects on the dynamics, power, and thermal efficiency of a pair of weakly coupled LTD Stirling engines and to formulate the maximum thermal efficiency of the coupled system in the quasilinear response regime. We will provide a model of a pair of weakly coupled LTD Stirling engines and investigate the coupling effects on the dynamics through numerical experiments. After that, we will provide a theoretical analysis of the effects of the weak coupling on power and thermal efficiency. By generalizing thermodynamic fluxes and forces and their quasilinear relations for engines under weak coupling, we show that the coupling improves the power exerted against the load torques and the thermal efficiency. We further show that their maximum values are achieved when the engines are synchronized. Finally, we formulate the load torques that achieve the maximum thermal efficiency of the coupled system. 

\section{Model}

\begin{figure*}[htbp]
\begin{center}
\includegraphics[width=150mm]{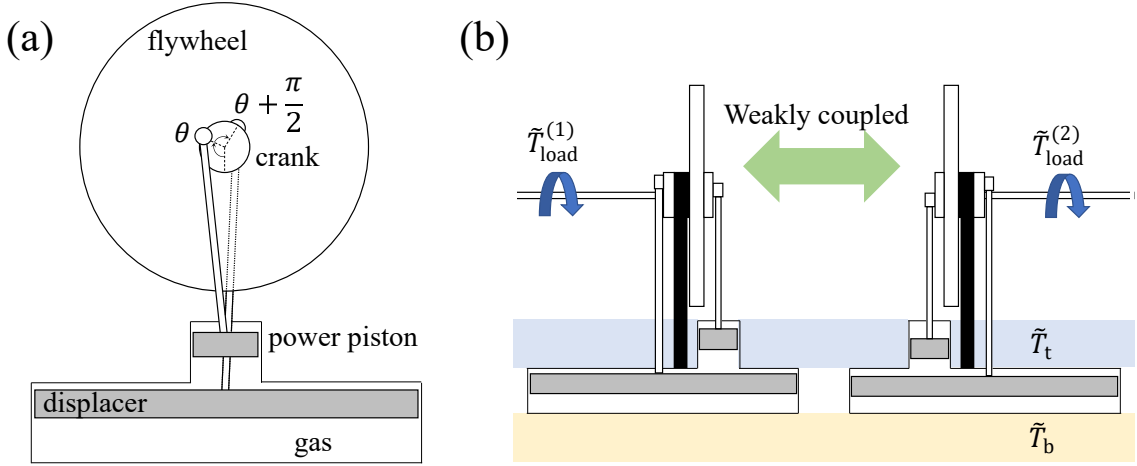}
\end{center}
\caption{(a) Front view of an LTD Stirling engine. (b) Side view of a pair of weakly coupled LTD Stirling engines with different load torques acting on the cranks.
The gases confined to the cylinders are in contact with the bottom and top heat reservoirs.
}\label{fig.1}
\end{figure*}

We consider a pair of weakly coupled LTD Stirling engines with the same parameters except for the load torques $\tilde{T}_{\rm load}^{(1)}$ and $\tilde{T}_{\rm load}^{(2)}$ acting on the cranks (Fig. \ref{fig.1}). Heat reservoirs at temperatures $\tilde{T}_{\rm b}$ and $\tilde{T}_{\rm t}$ ($\tilde{T}_{\rm b}>\tilde{T}_{\rm t}$) are attached to the bottom and top surfaces of the large cylinders of the engines respectively, and we define the temperature difference $\Delta\tilde{T}\equiv\tilde{T}_{\rm b}-\tilde{T}_{\rm t}$ for later use. The temperature difference $\Delta\tilde{T}$ and load torque $\tilde{T}_{\rm load}^{(i)}$ ($i=1,2$) are assumed to be sufficiently small. A nondimensionalized minimal model of a single LTD Stirling engine has been proposed in \cite{izumida2018nonlinear} with the following form: 
\begin{subequations}
\begin{equation}\label{single_dynamics0}
\frac{d\theta}{dt}=\omega,
\end{equation}
\begin{equation}\label{single_dynamics}
\frac{d\omega}{dt}=\sigma\left(\frac{T(\theta,\omega)}{V(\theta)}-P_{\rm air}\right)\sin\theta-\Gamma\omega-T_{\rm load},
\end{equation}
\end{subequations}
where $\theta$ is the phase angle of the crank connected to the power piston; $\sigma$ is a positive constant determined by the surface areas of the large and small cylinders; $V(\theta)=2+\sigma(1-\cos\theta)$ and $T(\theta,\omega)=T_{\rm eff}(\theta)/\left(1+\frac{\sigma\sin\theta\omega}{GV(\theta)}\right)$ represent the volume and temperature of the gas confined to the cylinders, respectively; $T_{\rm eff}(\theta)=1+\frac{\sin\theta}{2}\Delta T$ is the effective temperature of the heat reservoirs that periodically changes depending on the phase angle; $G$ is the thermal conductance associated with the heat transfer between the gas and the surface of the large cylinder; $P_{\rm air}$ is the atmospheric pressure acting on the power piston, and $\Gamma$ is the friction coefficient associated with the power piston. All the variables and parameters that are off-tilde represent dimensionless quantities.
The minimal model was obtained by assuming that the heat fluxes from the bottom and top surfaces of the large cylinder obey the Fourier law $J_{Q_{\rm m}}=G_{\rm m}(\theta)(T_{\rm m}-T(\theta,\omega))$, where $G_{\rm m}(\theta)$ with ${\rm m}={\rm b}$ (or t) represents the effective thermal conductance between the gas and the bottom (or top) heat reservoir. It was also assumed that $G_{\rm m}(\theta)\equiv G\chi_{\rm m}(\theta)$, where $\chi_{\rm m}(\theta)\hspace{1mm} (0\leq\chi_{\rm m}(\theta)\leq1)$ is a function that controls the coupling between the gas and the bottom or top heat reservoir, given as $\chi_{\rm b}(\theta)=\frac{1}{2}(1+\sin\theta)$ and $\chi_{\rm t}(\theta)=\frac{1}{2}(1-\sin\theta)$ \cite{izumida2020quasilinear}.
The dynamical equations describe the engines as coupled nonlinear pendulums, where the first term on the RHS of Eq. (\ref{single_dynamics}) represents the driven force due to the temperature difference. Since it has been experimentally demonstrated that the minimal model (\ref{single_dynamics0})-(\ref{single_dynamics}) explains the essential characteristics of a real LTD Stirling engine \cite{toyabe2020experimental}, we generalize the above minimal model by adding a coupling term to describe the dynamics of a pair of weakly-coupled LTD Stirling engines $i$ and $j$ ($i,j\in\{1,2\}, i\neq j$):
\begin{subequations}
\begin{equation}\label{dynamics0}
\frac{d\theta_i}{dt}=\omega_i,
\end{equation}
\begin{equation}\label{dynamics}
\frac{d\omega_i}{dt}=\sigma\left(\frac{T(\theta_i,\omega_i)}{V(\theta_i)}-P_{\rm air}\right)\sin\theta_i-\Gamma\omega_i-T_{\rm load}^{(i)}-K\sin(\theta_i-\theta_j).
\end{equation}
\end{subequations}
The last term in Eq. (\ref{dynamics}) represents the coupling with $K>0$ being the coupling strength. Note that the coupling should be anti-symmetric according to the action-reaction law and is chosen to be a sine function for simplicity. 

\section{Coupling effects on the dynamics}

\begin{figure*}[htbp]
\begin{center}
\includegraphics[width=150mm]{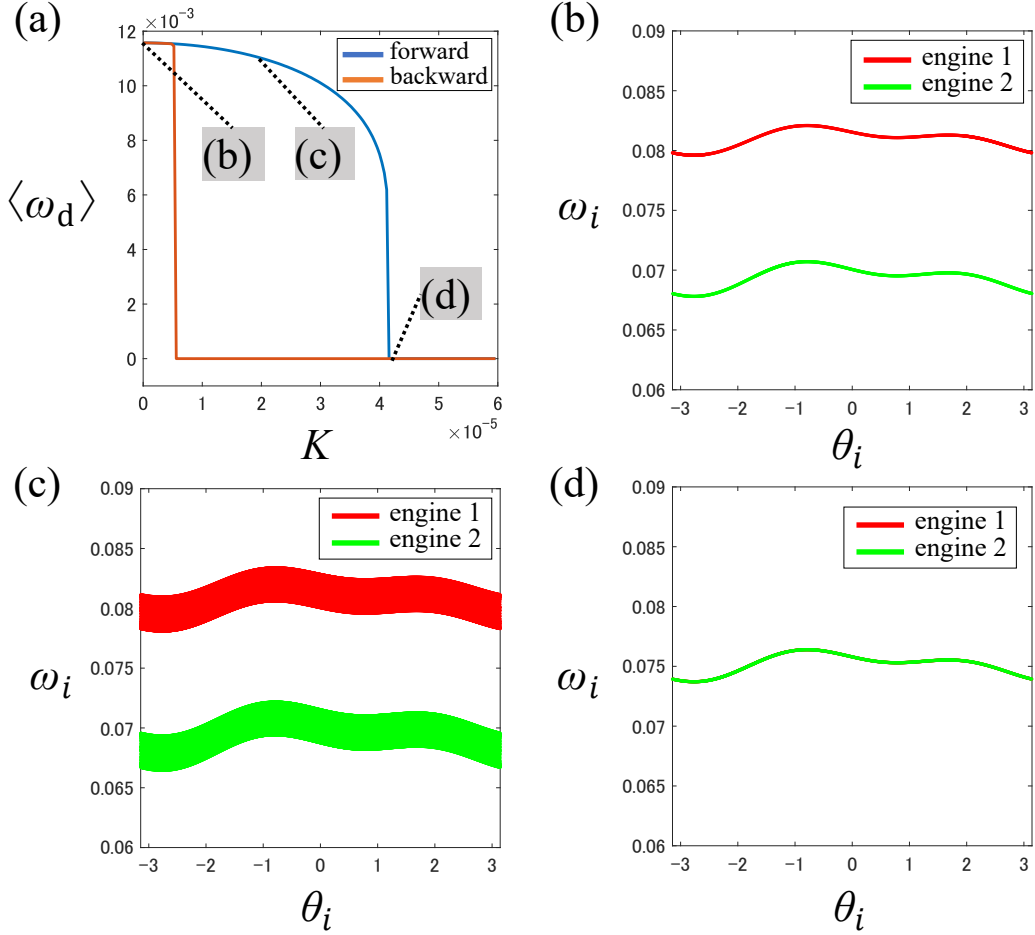}
\end{center}
\caption{(a) Dependence relation between the effective frequency difference 
$\langle\omega_{\rm d}\rangle$
and coupling strength $K$. 
In the forward (backward) processes, $K$ is increased (decreased) between $0$ and $6.0\times 10^{-5}$ with a step of $4.0\times 10^{-7}$. We can confirm that the transition point $K_{\rm fd}$ in the forward process is different from the transition point $K_{\rm bd}$ in the backward process.
Typical trajectories in the forward process are shown in the $(\theta_i, \omega_i)$ plane: (b) $K=0$. The trajectory of each engine is a limit cycle that circles the phase cylinder. (c) $K=1.60\times10^{-5}$. The trajectory of each engine evolves quasi-periodically and vibrates repeatedly in the vertical direction around a certain value. (d) $K=4.1522\times10^{-5}$. The trajectory of each engine is a limit cycle whose periods are the same, i.e., $\langle\omega_1\rangle=\langle\omega_2\rangle$, indicating that the two engines are synchronized. Other parameters are set as follows in all subsequent numerical experiments: $\sigma=0.02, \hspace{1mm}p_{\rm air}=\frac{1}{V\left(\frac{\pi}{4}\right)}\approx0.49854, \hspace{1mm}G=1.5, \hspace{1mm}\Gamma=0.001, \hspace{1mm}\Delta T=1/29.3, \hspace{1mm}T_{\rm load}^{(1)}=8.5324\times10^{-7}$, and $T_{\rm load}^{(2)}=1.2799\times10^{-5}$. The values of the parameters other than the load torques are set as the same as Ref. \cite{izumida2020quasilinear}, corresponding to the situation where a standard LTD Stirling engine is placed at a temperature difference of a few degrees. The load torques $T_{\rm load}^{(1)}$ and $T_{\rm load}^{(2)}$ are chosen so that $\omega_i^{\rm var}\ll\Delta\omega_{\rm n}\ll\omega_{\rm n}^{(i)}$. Similar graphs can be obtained with a different set of parameter values that satisfy the above conditions.
}
\label{fig.2}
\end{figure*}

To evaluate the degree of synchronization caused by the coupling, we introduce the effective frequency as
\begin{align}
\langle\omega_i\rangle=\lim_{\tau\to\infty}\frac{1}{\tau}\int_0^\tau\omega_idt,
\end{align}
where $\langle...\rangle\equiv\lim_{\tau\to0}\frac{1}{\tau}\int_0^\tau...dt$ denotes a long-time average and is reduced to the average over one period for engines in periodic motion. For $K=0$, the engines are adjusted to be in the quasilinear response regime \cite{izumida2020quasilinear} so that they rotate autonomously in a self-sustained manner. The phase space of engine $i$ is a set of ordered pairs $\{(\theta_i, \omega_i):\theta_i\in[-\pi, \pi), \omega_i\in\mathbb{R}\}$ which is a one-dimensional cylinder $\mathbb T \times \mathbb R$, and the rotational motion is described by a limit cycle that circles the surface of the cylinder.
The limit cycle for $K=0$ is referred to as the unperturbed limit cycle.
Since each engine behaves as a limit-cycle oscillator, we can define the natural frequencies of the two engines and denote them by $\omega_{\rm n}^{(i)}$ where $i=1, 2$, i.e., $\omega_{\rm n}^{(i)}=\langle\omega_i\rangle$ for $K=0$. We assume that the natural frequency difference $\Delta\omega_{\rm n}\equiv|\omega_{\rm n}^{(1)}-\omega_{\rm n}^{(2)}|$ is sufficiently small compared to $\omega_{\rm n}^{(i)}$, but sufficiently larger than the $\omega_i$-directional amplitude of the unperturbed limit cycle in the phase space, i.e., $\omega_i^{\rm var}\ll\Delta\omega_{\rm n}\ll\omega_{\rm n}^{(i)}$, where $\omega_i^{\rm var}$ is given by $\omega_i^{\rm var}\equiv\max_{s_i\in\Omega_i^{\rm UNP}}|s_i-\omega_{\rm n}^{(i)}|$ and $\Omega_i^{\rm UNP}$ denote the set of $\omega_i$-components of the points on the unperturbed limit cycle of engine $i$. For $K>0$, the two engines are coupled with each other and synchronization occurs for a sufficiently large coupling strength. Figure \ref{fig.2} shows the dependence relation between the effective frequency difference $\langle\omega_{\rm d}\rangle\equiv\langle\omega_1\rangle-\langle\omega_2\rangle$ and the coupling strength $K$, where the forward (backward) process corresponds to the situation in which the value of $K$ is increased (decreased). Typical trajectories in the $(\theta_i, \omega_i)$ plane are also shown in the same figure.

The bifurcation diagram in Fig. \ref{fig.2} reminds us of the dynamics of a driven pendulum \cite{Strogatz} or a two-node power grid model consisting of one generator and one consumer \cite{manik2014supply,rohden2012self}, where a homoclinic and saddle-node bifurcation for fixed points occur in the forward and backward processes respectively. Our numerical analysis indicates that similar bifurcations occur in weakly coupled LTD Stirling engines.
Particularly, when considering the differential system of (\ref{dynamics0})-(\ref{dynamics}), a homoclinic bifurcation due to the annihilation of a quasi-periodic trajectory is thought to occur in the forward process as a result of the collision of this quasi-periodic trajectory with a saddle limit cycle corresponding to an unstable synchronous state, and a saddle-node bifurcation is thought to occur in the backward process due to the collision of a stable limit cycle corresponding to a stable synchronous state and an unstable limit cycle corresponding to an unstable synchronous state (See Appendix A for details).

\section{Coupling effects on the power and thermal efficiency}
The power and thermal efficiency of a single LTD Stirling engine in a quasilinear response regime have been derived in \cite{izumida2020quasilinear}. Before discussing the coupling effects on the power and thermal efficiency of the total system, we generalize the thermodynamic fluxes and forces as well as their quasilinear relations for engines under weak coupling.

The instantaneous power $P^{(i)}$ produced by engine $i$ is given by
\begin{align}
P^{(i)}=\frac{d}{dt}\left(\frac{1}{2}\omega_i^2\right)+P_{\rm air}\frac{dV}{dt}+P_{\rm load}^{(i)}+P_{\rm fric}^{(i)}+P_{K}^{(i)}
\end{align}
where $\frac{d}{dt}\left(\frac{1}{2}\omega_i^2\right)$ is the change rate of rotational energy, $P_{\rm air}\frac{dV}{dt}$ is the power that is carried out against the atmospheric pressure, $P_{\rm load}^{(i)}\equiv T_{\rm load}^{(i)}\omega_i$ is the power that is carried out against the load torque, $P_{\rm fric}^{(i)}\equiv\Gamma\omega_i^2$ is the power that is carried out against the friction torque, and $P_{K}^{(i)}\equiv K\sin(\theta_i-\theta_j)\omega_i$ is the power due to the weak coupling. The power that is carried out against the load torque $P_{\rm load}^{(i)}$ is referred to as the brake power \cite{medina2014quasi} made by engine $i$. Since $\bigl\langle\frac{d}{dt}\left(\frac{1}{2}\omega_i^2\right)\bigr\rangle=\bigl\langle P_{\rm air}\frac{dV}{dt}\bigr\rangle=0$, the time-averaged power of engine $i$ is obtained as $\langle P^{(i)}\rangle=\bigl\langle P_{\rm load}^{(i)}\rangle+\bigl\langle P_{\rm fric}^{(i)}\bigr\rangle+\bigl\langle P_{K}^{(i)}\bigr\rangle$. It should be noted that for the system in quasi-periodic motion, the trajectory is not closed, so the long-time average can not be reduced to the average over an oscillation period. Given that the coupling is sufficiently weak and each engine is in the quasilinear response regime when there is no coupling, $\omega_i$ can be approximated to be the effective frequency $\langle\omega_i\rangle$ when considering $\bigl\langle P_{K}^{(i)}\bigr\rangle$, i.e.,
$\bigl\langle P_{K}^{(i)}\bigr\rangle$ can be approximated as $\bigl\langle P_{K}^{(i)}\bigr\rangle\approx K\langle\sin(\theta_i-\theta_j)\rangle\langle\omega_i\rangle$.
Since the time-averaged changes in the entropy and the internal energy of the gas confined to the cylinder are zero, the time-averaged entropy production rate of the total thermodynamic system $\bigl\langle\frac{d\sigma}{dt}\bigr\rangle$ is the sum of the time-averaged entropy change rates of the two heat baths,
which is calculated as
\begin{align}
\biggl\langle\frac{d\sigma}{dt}\biggr\rangle&=\sum_{i=1}^2
\left[ -\frac{\bigl\langle J_{Q_{\rm b}}^{(i)}\bigr\rangle}{T_{\rm b}}-\frac{\bigl\langle J_{Q_{\rm t}}^{(i)}\bigr\rangle-\bigl\langle P_{\rm fric}^{(i)}\bigr\rangle}{T_{\rm t}}
\right]\\
&=\sum_{i=1}^2 \left[ -\frac{\langle P_{\rm load}^{(i)}\rangle+\bigl\langle P_{K}^{(i)}\bigr\rangle}{T_{\rm t}} 
+\bigl\langle J_{Q_{\rm b}}^{(i)}\bigr\rangle\left(\frac{1}{T_{\rm t}}-\frac{1}{T_{\rm b}}\right)\right]\label{entropy_conservation}\\
&\approx\sum_{i=1}^2\left[-\langle\omega_i\rangle T_{{\rm load}}^{(i)}+\bigl\langle J_{Q_{\rm b}}^{(i)}\bigr\rangle\Delta T\right]
+K\langle\sin(\theta_1-\theta_2)\rangle\left(\langle\omega_2\rangle-\langle\omega_1\rangle\right)\label{entropy_approx}\\
&=\langle\omega_{\rm m}\rangle\left(-T_{{\rm load}}^{(1)}-T_{{\rm load}}^{(2)}\right)+\langle\omega_{\rm d}\rangle\left[-K\langle\sin(\theta_1-\theta_2)\rangle-\frac{1}{2}\left(T_{{\rm load}}^{(1)}-T_{{\rm load}}^{(2)}\right)\right]+\left(\bigl\langle J_{Q_{\rm b}}^{(1)}\bigr\rangle+\bigl\langle J_{Q_{\rm b}}^{(2)}\bigr\rangle\right)\Delta T\label{entropy_final},
\end{align}
where we have used the energy conservation law $\bigl\langle J_{Q_{\rm b}}^{(i)}\bigr\rangle+\bigl\langle J_{Q_{\rm t}}^{(i)}\bigr\rangle=\bigl\langle P_{\rm load}^{(i)}\rangle+\bigl\langle P_{\rm fric}^{(i)}\bigr\rangle+\bigl\langle P_{K}^{(i)}\bigr\rangle$ in Eq. (\ref{entropy_conservation}) and approximated $T_{\rm b}$ and $T_{\rm t}$ as their mean value in Eq. (\ref{entropy_approx}), which equals $1$ for the nondimensionalized case. Here, $\langle\omega_{\rm m}\rangle\equiv\frac{1}{2}\left(\langle\omega_1\rangle+\langle\omega_2\rangle\right)$ is the mean effective frequency, $\langle\omega_{\rm d}\rangle=\langle\omega_1\rangle-\langle\omega_2\rangle$ is the effective frequency difference, and $\bigl\langle J_{Q_{\rm b}}^{(1)}\bigr\rangle+\bigl\langle J_{Q_{\rm b}}^{(2)}\bigr\rangle$ is the total heat flux from the high-temperature heat reservoir. 

Equation (\ref{entropy_final}) suggests that $-T_{{\rm load}}^{(1)}-T_{{\rm load}}^{(2)}$, $-K\langle\sin(\theta_1-\theta_2)\rangle-\frac{1}{2}\left(T_{{\rm load}}^{(1)}-T_{{\rm load}}^{(2)}\right)$, and $\Delta T$ can be considered as thermodynamic forces with conjugate fluxes $\langle\omega_{\rm m}\rangle$, $\langle\omega_{\rm d}\rangle$, and $\bigl\langle J_{Q_{\rm b}}^{(1)}\bigr\rangle+\bigl\langle J_{Q_{\rm b}}^{(2)}\bigr\rangle$ under appropriate conditions, for which the quasilinear relations are obtained as follows (See Appendix B for details): 
\begin{equation}\label{quasilinear}
\left[
  \begin{array}{c}
      \bigl\langle\omega_{\rm m}\bigr\rangle \\
      \langle\omega_{\rm d}\rangle \\
      \bigl\langle J_{Q_{\rm b}}^{(1)}\bigr\rangle+\bigl\langle J_{Q_{\rm b}}^{(2)}\bigr\rangle \\
    \end{array}
  \right]
  \approx
  \left[
  \begin{array}{ccc}
      \frac{1}{2}L_{11} & 0 & L_{12}  \\
      0 & 2L_{11} & 0  \\
      L_{12} & 0 & 2L_{22}  \\
    \end{array}
  \right]
  \left[
  \begin{array}{c}
      -T_{{\rm load}}^{(1)}-T_{{\rm load}}^{(2)} \\
      -K\langle\sin(\theta_1-\theta_2)\rangle-\frac{1}{2}\left(T_{{\rm load}}^{(1)}-T_{{\rm load}}^{(2)}\right) \\
      \Delta T \\
  \end{array}
  \right].
\end{equation}
Here, $L_{11}$, \hspace{1mm}$L_{12}$, \hspace{1mm}$L_{21}$, and \hspace{1mm}$L_{22}$ correspond to the quasilinear response coefficients of a single engine in the non-coupling case \cite{izumida2020quasilinear}:
\begin{equation}
L_{11}=\frac{1}{\Gamma+\frac{\sigma^2}{G}\Bigl\langle\frac{\sin^2\theta}{V^2(\theta)}\Bigr\rangle_\theta},
\end{equation}
\begin{equation}
L_{12}=L_{21}=\frac{\frac{\sigma}{2}\Bigl\langle\frac{\sin^2\theta}{V(\theta)}\Bigr\rangle_\theta}{\Gamma+\frac{\sigma^2}{G}\Bigl\langle\frac{\sin^2\theta}{V^2(\theta)}\Bigr\rangle_\theta},
\end{equation}
\begin{equation}
L_{22}=\frac{G}{8}+\frac{\frac{\sigma^2}{4}\Bigl\langle\frac{\sin^2\theta}{V(\theta)}\Bigr\rangle_\theta^2}{\Gamma+\frac{\sigma^2}{G}\Bigl\langle\frac{\sin^2\theta}{V^2(\theta)}\Bigr\rangle_\theta},
\end{equation}
where $\langle...\rangle_\theta\equiv\frac{1}{2\pi}\int_0^{2\pi}...d\theta$ denotes a phase average.

We now consider the coupling effects on the averaged brake power $\langle P_{\rm load}\rangle\equiv T_{{\rm load}}^{(1)}\langle\omega_1\rangle+T_{{\rm load}}^{(2)}\langle\omega_2\rangle$ and thermal efficiency $\eta\equiv\frac{\langle P_{\rm load}\rangle}{\bigl\langle J_{Q_{\rm b}}^{(1)}\bigr\rangle+\bigl\langle J_{Q_{\rm b}}^{(2)}\bigr\rangle}$ by using the generalized quasilinear relations (\ref{quasilinear}) between thermodynamic fluxes and forces. To that end, we rewrite the averaged brake power $\langle P_{\rm load}\rangle$ in the following form:
\begin{align}
\langle P_{\rm load}\rangle
&=T_{{\rm load}}^{(1)}\left(\bigl\langle\omega_{\rm m}\bigr\rangle+\frac{1}{2}\langle\omega_{\rm d}\rangle\right)+T_{{\rm load}}^{(2)}\left(\bigl\langle\omega_{\rm m}\bigr\rangle-\frac{1}{2}\langle\omega_{\rm d}\rangle\right)\\
&=\bigl\langle\omega_{\rm m}\bigr\rangle\left(T_{{\rm load}}^{(1)}+T_{{\rm load}}^{(2)}\right)+\frac{1}{2}\langle\omega_{\rm d}\rangle\left(T_{{\rm load}}^{(1)}-T_{{\rm load}}^{(2)}\right)\\
&=\langle P_{\rm m}\rangle+\langle P_{\rm rel}\rangle.
\end{align}
Here, $\langle P_{\rm m}\rangle\equiv\bigl\langle\omega_{\rm m}\bigr\rangle\left(T_{{\rm load}}^{(1)}+T_{{\rm load}}^{(2)}\right)$ denotes the power owing to the motion of the mean angle and $\langle P_{\rm rel}\rangle\equiv\frac{1}{2}\langle\omega_{\rm d}\rangle\left(T_{{\rm load}}^{(1)}-T_{{\rm load}}^{(2)}\right)$ denotes the power owing to the relative motion. Since $\bigl\langle\omega_{\rm m}\bigr\rangle$ is independent of the coupling from Eq. (\ref{quasilinear}), we need only consider the coupling effects on $\langle P_{\rm rel}\rangle$. Without loss of generality, we assume $T_{{\rm load}}^{(1)}<T_{{\rm load}}^{(2)}$, in which case the value of $\langle\omega_{\rm d}\rangle$ decreases due to the effect of the coupling strength $K$ in both forward and backward processes, as was shown in Fig. \ref{fig.2}. This leads to the fact that $\langle P_{\rm rel}\rangle$ is an increasing function of $K$, which means that the coupling improves the averaged power. To see the coupling effect on the thermal efficiency, we notice that the total heat flux from the high-temperature heat reservoir $\bigl\langle J_{Q_{\rm b}}^{(1)}\bigr\rangle+\bigl\langle J_{Q_{\rm b}}^{(2)}\bigr\rangle$ is independent of the coupling from Eq. (\ref{quasilinear}). This suggests that the coupling improves both the averaged brake power and the thermal efficiency given that different load torques act on the cranks, and their maximum values are achieved when the engines are synchronized.

To give a physical interpretation of the fact that a weak coupling improves $\langle P_{\rm rel}\rangle$, let us concentrate on $\langle\omega_{\rm d}\rangle$ in Eq. (\ref{quasilinear}). We find that $\langle\omega_{\rm d}\rangle$ in the non-coupling case is reduced to $-L_{11}\left(T_{{\rm load}}^{(1)}-T_{{\rm load}}^{(2)}\right)$, which means that $\langle P_{\rm rel}\rangle$ is generated by the synergy of the load torque difference and the relative motion due to the load torque difference when there is no coupling. In this case, $\bigl\langle P_{\rm rel}\bigr\rangle$ takes a negative value as long as $T_{{\rm load}}^{(1)}\neq T_{{\rm load}}^{(2)}$, and is a decreasing function of $|T_{{\rm load}}^{(1)}-T_{{\rm load}}^{(2)}|$. We thus conclude that $\bigl\langle P_{\rm rel}\bigr\rangle$ reduces the averaged brake power given a fixed sum of load torques. When there is a coupling added, $\langle P_{\rm rel}\rangle$ is obtained by
\begin{align}
\langle P_{\rm rel}\rangle
&\approx\frac{1}{2}\times2L_{11}\left[-K\langle\sin(\theta_1-\theta_2)\rangle-\frac{1}{2}\left(T_{{\rm load}}^{(1)}-T_{{\rm load}}^{(2)}\right)\right]\left(T_{{\rm load}}^{(1)}-T_{{\rm load}}^{(2)}\right)\label{P_rel_Delta_omega}\\
&=-\frac{1}{2}L_{11}\left(T_{{\rm load}}^{(1)}-T_{{\rm load}}^{(2)}\right)^2-L_{11}K\langle\sin(\theta_i-\theta_j)\rangle\left(T_{{\rm load}}^{(1)}-T_{{\rm load}}^{(2)}\right).
\end{align}
Here, $\langle\Delta P_{\rm rel}\rangle\equiv-L_{11}K\langle\sin(\theta_i-\theta_j)\rangle\left(T_{{\rm load}}^{(1)}-T_{{\rm load}}^{(2)}\right)$ represents the change of $\langle P_{\rm rel}\rangle$ due to the coupling, which takes a positive value as long as $T_{{\rm load}}^{(1)}\neq T_{{\rm load}}^{(2)}$. This suggests that the coupling improves the averaged brake power. From Eq. (\ref{P_rel_Delta_omega}), we find that the increase in averaged brake power is due to the suppression effect of coupling on relative motion caused by the load torque difference. The averaged brake power takes the maximum value when $K\langle\sin(\theta_i-\theta_j)\rangle$ reaches $-\frac{1}{2}\left(T_{{\rm load}}^{(1)}-T_{{\rm load}}^{(2)}\right)$, in which case $\bigl\langle\omega_{\rm d}\bigr\rangle=0$, meaning that the engines are synchronized.

\begin{figure*}[htbp]
\begin{center}
\includegraphics[width=170mm]{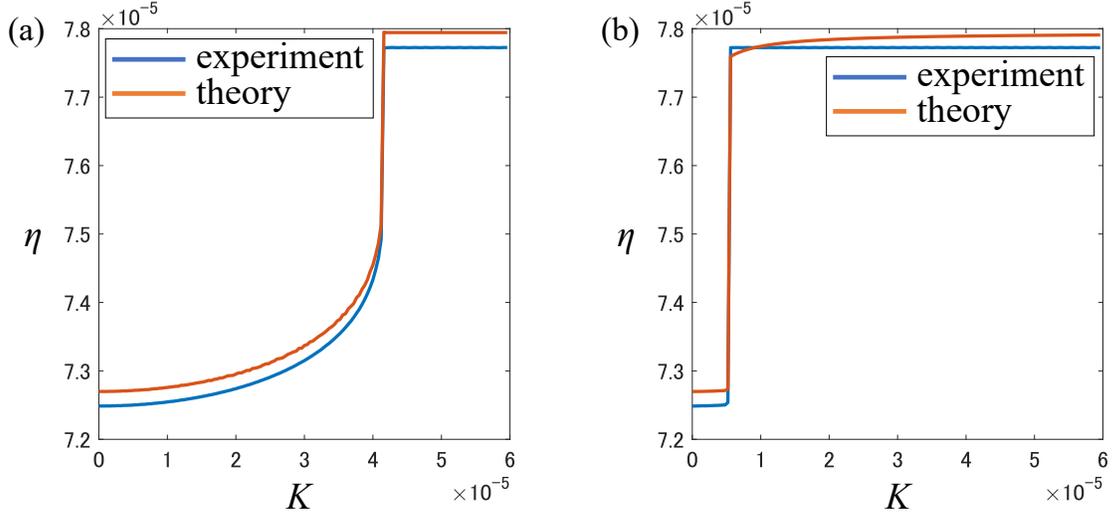}
\end{center}
\caption{Dependence relation between the thermal efficiency and the coupling strength for (a) forward process and (b) backward process. $K$ is increased from $0$ to $6.0\times 10^{-5}$ in increments of $4.0\times 10^{-7}$ in the forward process and decreased in the same way in the backward process. In the current case, we can confirm that the coupling has increased the thermal efficiency of the total system by about 7\%.}\label{fig.3}
\end{figure*}

Figure \ref{fig.3} shows the dependence relation between the thermal efficiency and the coupling strength in the forward and backward processes. The blue line is obtained by numerical experiment, while the orange line is obtained by approximate calculation using the quasilinear relations between thermodynamics fluxes and forces. Since it is difficult to calculate $K\langle\sin(\theta_1-\theta_2)\rangle$ analytically, we used numerical values of it in the approximate calculation. We can see some gap between experimental and theoretical values, which is caused by neglecting higher-order terms and by the averaging approximation made in the derivation of Eq. (\ref{quasilinear}). We also find that the dependence of the thermal efficiency on the coupling strength is characterized by a hysteresis as in the case of the frequency difference in Fig. \ref{fig.2} (a). This is because the thermal efficiency depends on the effective frequencies of the two engines. Such a hysteresis structure facilitates the robustness of maintaining maximum thermal efficiency.

We have confirmed that coupling can improve the averaged brake power and thermal efficiency. Since the total load torque determines the power and thermal efficiency given a fixed coupling strength, it is important to investigate the total load torque that achieves the maximum values of them for synchronized engines. In this case, $\bigl\langle\omega_{\rm m}\bigr\rangle$ and $\bigl\langle\omega_{\rm d}\bigr\rangle$ are reduced to the synchronized frequency $\omega_{\rm s}$, and $0$, respectively,
indicating that $\omega_{\rm s}$ and $\bigl\langle J_{Q_{\rm b}}^{(1)}\bigr\rangle+\bigl\langle J_{Q_{\rm b}}^{(2)}\bigr\rangle$ are the only thermodynamic fluxes for the coupled system with conjugate forces $-T_{{\rm load}}^{(1)}-T_{{\rm load}}^{(2)}$ and $\Delta T$. The thermal efficiency is given by
\begin{equation}
\eta=\frac{\omega_{\rm s}\left(T_{{\rm load}}^{(1)}+T_{{\rm load}}^{(2)}\right)}{\bigl\langle J_{Q_{\rm b}}^{(1)}\bigr\rangle+\bigl\langle J_{Q_{\rm b}}^{(2)}\bigr\rangle},
\end{equation}
which is completely determined by the thermodynamic fluxes and forces. Therefore, the formulation of the maximum thermal efficiency of a single engine given in \cite{izumida2020quasilinear} is directly applicable to the present case. The maximum thermal efficiency $\eta_{\rm max}$ and the total load torque $T_{{\rm load}}^{(1)}+T_{{\rm load}}^{(2)}$ that achieves this maximum thermal efficiency are given by
\begin{equation}
\eta_{\rm max}=\frac{\left(1-\sqrt{1-q^2}\right)^2}{q^2}\eta_{\rm C},
\end{equation}
\begin{equation}
T_{{\rm load}}^{(1)}+T_{{\rm load}}^{(2)}=\frac{2L_{12}\Delta T\left(1-\sqrt{1-q^2}\right)}{q^2L_{11}},
\end{equation}
where $q\equiv\frac{L_{12}}{\sqrt{L_{11}L_{22}}}$ is the coupling-strength parameter and $\eta_{\rm C}\equiv 1-\frac{T_t}{T_b}$ is the Carnot efficiency, i.e., the maximum thermal efficiency that a heat engine may have operating between two heat reservoirs. We find that the coupling-strength parameter, as well as the maximum thermal efficiency, is of the same form as that of a single engine, while the total load torque that achieves the maximum thermal efficiency is twice as large as that of a single engine given a fixed temperature difference $\Delta T$. The load torques achieving the maximum power and the corresponding thermal efficiency \cite{van2005thermodynamic} can be discussed in the same way \cite{izumida2020quasilinear}.

\section{Discussion and Conclusions}
In this paper, we have considered a minimal dynamical-system model of weakly coupled LTD Stirling engines and analyzed the coupling effects on the dynamics, power, and thermal efficiency. We clarified the mechanism of different kinds of bifurcation in the forward and backward processes and generalized the thermodynamic fluxes and forces and their quasilinear relations when the weak coupling is incorporated. Based on the linear relations, we concluded that the coupling improves the power exerted against the load torque as well as the thermal efficiency and that their maximum values are achieved when two engines are synchronized. We formulated the maximum thermal efficiency given that the coupled engines are synchronized and found that the expression of the maximum thermal efficiency is given in the same form as that of a single engine. Although the thermal efficiency of LTD Stirling engines is low \cite{Aragon2013}, their great value lies in their ability to generate power with only a small temperature difference. In other words, unlike the large engines that are run in factories, they do not need fuel to generate power, and only a ubiquitous temperature difference (e.g., between air and ground) is needed for the engine to generate power. To achieve sufficient power for practical use, it is desirable to operate a large number of LTD Stirling engines, and synchronizing the engines may be an effective way to further improve power and thermal efficiency. This study discusses the effects of synchronizing two engines as the simplest case, but will be extended to the case of multiple (3 or more) engines in the future.

\section*{Acknowledgements}

This work was supported by JSPS KAKENHI Grant Numbers 21K12056, 19K03651, and 22K03450.

\vspace{2mm}
\section*{Appendix A: Bifurcation analysis of the coupled system}

\begin{figure*}[htbp]
\begin{center}
\includegraphics[width=180mm]{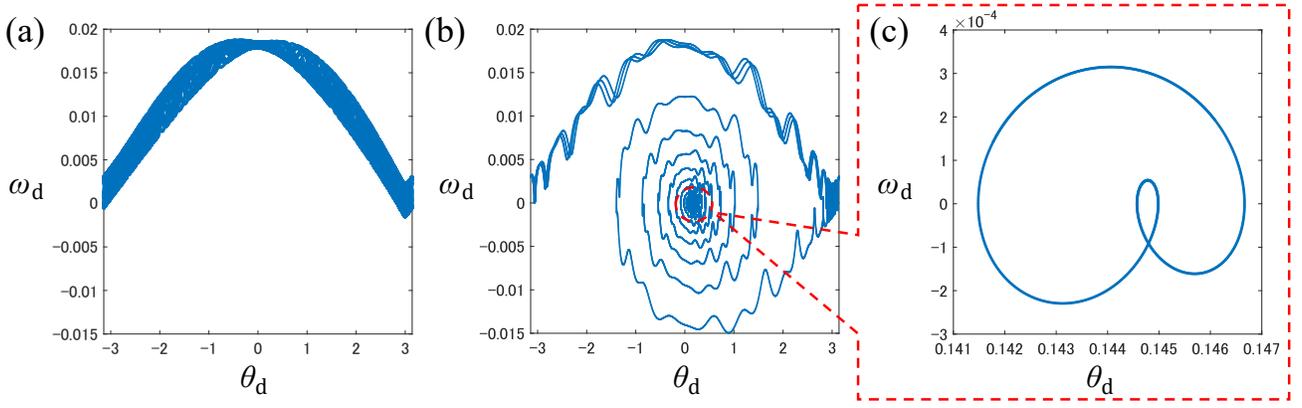}
\end{center}
\caption{Trajectories in the subspace $\{(\theta_{\rm d},\omega_{\rm d})\}$ in the forward process for different values of $K$. (a) $K$ is set slightly smaller than $K_{\rm fd}$ and the quasi-linear trajectory evolves much more slowly near $\theta_{\rm d}=3$ than elsewhere. (b) $K$ is set slightly larger than $K_{\rm fd}$. The trajectory evolves in the neighborhood of a homoclinic orbit starting and ending at a saddle limit cycle corresponding to an unstable synchronous state before converging to a stable limit cycle corresponding to a stable synchronous state. (c) Enlarged view of the stable synchronous state.
}
\label{fig.4}
\end{figure*}

To gain more insight into the bifurcations that occur when changing the coupling strength, we plot the trajectories in the subspace $\{(\theta_{\rm d},\omega_{\rm d})\}$ in the forward and backward processes before and after the bifurcations occur, where
$\theta_{\rm d} \equiv \theta_1 - \theta_2$ and
$\omega_{\rm d}\equiv\omega_1 - \omega_2$. In the forward process, when $K$ is raised to a value slightly less than the bifurcation point $K_{\rm fd}$, the quasi-periodic trajectory evolves much more slowly near $\theta_{\rm d}=3$ than elsewhere (See Fig. \ref{fig.4}. (a)); After the bifurcation, the trajectory converges to a stable limit cycle (See Fig. \ref{fig.4}. (b) and (c)). In the backward process, the stable limit cycle does not disappear until $K$ reaches another bifurcation point $K_{\rm bd}$; After the bifurcation, the stable limit cycle collapse and the trajectory converges to a quasi-periodic attractor circling the phase cylinder. These results suggest that a homoclinic bifurcation and a saddle-node bifurcation occurs in the forward and backward process respectively: in the forward process, the quasi-periodic trajectory evolves in the neighborhood of the stable and unstable manifolds of a saddle limit cycle corresponding to an unstable synchronous state when $K$ is slightly smaller than $K_{\rm fd}$, and converges to a stable limit cycle corresponding to a stable synchronous state after the bifurcation occurs; in the backward process, a saddle-node bifurcation occurs due to the collision of the stable and unstable synchronous states.

\section*{Appendix B: Derivation of Eq. (9)}

We derive the quasilinear relations between thermodynamic fluxes and forces given by Eq. (\ref{quasilinear}). To that end, we average both sides of Eq. (\ref{dynamics}):
\begin{equation}\label{dynamics_averaging}
0=\sigma\Biggl\langle\left(\frac{T(\theta_i,\omega_i)}{V(\theta_i)}-P_{\rm air}\right)\sin\theta_i\Biggr\rangle-\Gamma\langle\omega_i\rangle-T_{\rm load}^{(i)}-K\langle\sin(\theta_i-\theta_j)\rangle,
\end{equation}
where we have used the fact that
\begin{equation}
\biggl\langle\frac{d\omega_i}{dt}\biggr\rangle=\lim_{\tau\to\infty}\frac{1}{\tau}\int_0^\tau\frac{d\omega_i}{dt}dt=\lim_{\tau\to\infty}\frac{1}{\tau}\left\{\omega_i(\tau)-\omega_i(0)\right\}=0
\end{equation}
since the trajectory remains in the neighborhood of the unperturbed limit cycle. By expanding $T(\theta_i,\omega_i)$ w.r.t. $\omega_i$ as
\begin{equation}
T(\theta_i,\omega_i)=T_{{\rm eff}}(\theta_i)-\frac{\sigma\sin\theta_i}{GV}\omega_i+\mathcal{O}(\Delta T\omega_i,\omega_i^2),
\end{equation}
the first term on the RHS of Eq. (\ref{dynamics_averaging}) can be obtained as
\begin{align}\label{dynamics_averaging_first}
\sigma\Biggl\langle\left(\frac{T(\theta_i,\omega_i)}{V(\theta_i)}-P_{\rm air}\right)\sin\theta_i\Biggr\rangle
=\sigma\Biggl\langle\left(\frac{T_{{\rm eff}}(\theta_i)}{V(\theta_i)}-\frac{\sigma\sin\theta_i}{GV^2(\theta_i)}\omega_i-P_{\rm air}\right)\sin\theta_i\Biggr\rangle+\bigl\langle\mathcal{O}(\Delta T\omega_i,\omega_i^2)\bigr\rangle.
\end{align}
Substituting Eq. (\ref{dynamics_averaging_first}) into Eq. (\ref{dynamics_averaging}), we obtain
\begin{align}\label{dynamics_averaging_rewritten}
\sigma\Biggl\langle\left(\frac{T_{{\rm eff}}(\theta_i)}{V(\theta_i)}-P_{\rm air}\right)\sin\theta_i\Biggr\rangle-\frac{\sigma^2}{G}\Biggl\langle\frac{\sin^2\theta_i}{V^2(\theta_i)}\omega_i\Biggr\rangle
-\Gamma\langle\omega_i\rangle-T_{{\rm load}}^{(i)}-K\langle\sin(\theta_i-\theta_j)\rangle+\bigl\langle\mathcal{O}(\Delta T\omega_i,\omega_i^2)\bigr\rangle=0.
\end{align}
Let $\tau_k^{(i)}$ be the time required for $\theta_i$ to increase from $\theta_i(0)+2(k-1)\pi$ to $\theta_i(0)+2k\pi$. The first and second terms on the LHS in Eq. (\ref{dynamics_averaging_rewritten}) can then be calculated as follows:
\begin{align}
&\sigma\Biggl\langle\left(\frac{T_{{\rm eff}}(\theta_i)}{V(\theta_i)}-P_{\rm air}\right)\sin\theta_i\Biggr\rangle\notag\\
&=\sigma\left(\lim_{\tau\to\infty}\frac{1}{\tau}\int_0^{\tau}\left(\frac{T_{{\rm eff}}(\theta_i)}{V(\theta_i)}-P_{\rm air}\right)\sin\theta_idt\right)\\
&=\sigma\left(\lim_{N\to\infty}\frac{1}{\bigl\langle\omega_i\bigr\rangle\sum_{k=1}^N\tau_k^{(i)}}\int_0^{\sum_{k=1}^N\tau_k^{(i)}}\left(\frac{T_{{\rm eff}}(\theta_i)}{V(\theta_i)}-P_{\rm air}\right)\sin\theta_i\langle\omega_i\rangle dt\right)\\
&=\sigma\left(\lim_{N\to\infty}\frac{1}{2\pi N}\int_0^{2\pi N}\left(\frac{T_{{\rm eff}}(\theta)}{V(\theta)}-P_{\rm air}\right)\sin\theta d\theta\right)+\bigl\langle\mathcal{O}(\Delta\theta_i)\bigr\rangle\\
&=\frac{\sigma}{2}\Biggl\langle\frac{\sin^2\theta}{V(\theta)}\Biggr\rangle_\theta\Delta T+\bigl\langle\mathcal{O}(\Delta\theta_i)\bigr\rangle\label{dynamics_averaging_rewritten_first},
\end{align}
\begin{align}
\frac{\sigma^2}{G}\Biggl\langle\frac{\sin^2\theta_i}{V^2(\theta_i)}\omega_i\Biggr\rangle
&=
\frac{\sigma^2}{G}\left(\lim_{\tau\to\infty}\frac{1}{\tau}\int_0^{\tau}\frac{\sin^2\theta_i}{V^2(\theta_i)}\omega_idt\right)\\
&=
\frac{\sigma^2}{G}\left(\lim_{N\to\infty}\frac{1}{\sum_{k=1}^N\tau_k^{(i)}}\int_0^{2\pi N}\frac{\sin^2\theta}{V^2(\theta)}d\theta\right)\\
&=
\frac{\sigma^2}{G}\left(\frac{1}{2\pi}\int_0^{2\pi }\frac{\sin^2\theta}{V^2(\theta)}d\theta\right)\left(\lim_{N\to\infty}\frac{2\pi N}{\sum_{k=1}^N\tau_k^{(i)}}\right)\\
&=
\frac{\sigma^2}{G}\Biggl\langle\frac{\sin^2\theta}{V^2(\theta)}\Biggr\rangle_\theta\langle\omega_i\rangle\label{dynamics_averaging_rewritten_second},
\end{align}
where $\Delta\theta_i(t)\equiv\int_0^t\omega_i(\tau)d\tau-\langle\omega_i\rangle t$ and $\bigl\langle\mathcal{O}(\Delta\theta_i)\bigr\rangle$ denotes the error due to approximating $\omega_i$ to $\langle\omega_i\rangle$ in the calculation of the long-time average. Substituting Eq. (\ref{dynamics_averaging_rewritten_first}) and Eq. (\ref{dynamics_averaging_rewritten_second}) into Eq. (\ref{dynamics_averaging_rewritten}), we obtain
\begin{align}\label{omega_m&d_1}
\left(\Gamma+\frac{\sigma^2}{G}\Biggl\langle\frac{\sin^2\theta}{V^2(\theta)}\Biggr\rangle_\theta\right)\langle\omega_{\rm m}\rangle=\left[\frac{\sigma}{2}\Biggl\langle\frac{\sin^2\theta}{V(\theta)}\Biggr\rangle_\theta\Delta T-\frac{1}{2}\left(T_{\rm load}^{(1)}+T_{\rm load}^{(2)}\right)\right]+\bigl\langle\mathcal{O}(\Delta T\omega_i,\omega_i^2,\Delta\theta_i)\bigr\rangle,
\end{align}
\begin{align}\label{omega_m&d_2}
\left(\Gamma+\frac{\sigma^2}{G}\Biggl\langle\frac{\sin^2\theta}{V^2(\theta)}\Biggr\rangle_\theta\right)\langle\omega_{\rm d}\rangle=\left[-2K\langle\sin\theta_{\rm d}\rangle-\left(T_{\rm load}^{(1)}-T_{\rm load}^{(2)}\right)\right]+\bigl\langle\mathcal{O}(\Delta T\omega_i,\omega_i^2,\Delta\theta_i)\bigr\rangle,
\end{align}
where $\langle\omega_{\rm m}\rangle=\frac{1}{2}\left(\langle\omega_1\rangle+\langle\omega_2\rangle\right)$ is the mean effective frequency, and $\langle\omega_{\rm d}\rangle=\langle\omega_1\rangle-\langle\omega_2\rangle$ is the effective frequency difference. By neglecting the higher order terms in Eqs. (\ref{omega_m&d_1}) and (\ref{omega_m&d_2}), we can obtain the quasi-linear relations w.r.t. $\langle\omega_{\rm m}\rangle$ and $\langle\omega_{\rm d}\rangle$ in Eqs. (\ref{quasilinear}), and the effective frequency $\langle\omega_i\rangle$ of each engine:
\begin{equation}\label{omega_average}
\langle\omega_i\rangle\approx\frac{-T_{{\rm load}}^{(i)}-K\langle\sin(\theta_i-\theta_j)\rangle+\frac{\sigma}{2}\Bigl\langle\frac{\sin^2\theta}{V(\theta)}\Bigr\rangle_\theta\Delta T}{\Gamma+\frac{\sigma^2}{G}\Bigl\langle\frac{\sin^2\theta}{V^2(\theta)}\Bigr\rangle_\theta}.
\end{equation}
If $K$ is large enough so that the two engines are synchronized, the synchronized frequency $\omega_{\rm s}$ can be obtained as
\begin{equation}
\omega_{\rm s}\approx\frac{-\frac{1}{2}\left(T_{{\rm load}}^{(1)}+T_{{\rm load}}^{(2)}\right)+\frac{\sigma}{2}\Bigl\langle\frac{\sin^2\theta}{V(\theta)}\Bigr\rangle_\theta\Delta T}{\Gamma+\frac{\sigma^2}{G}\Bigl\langle\frac{\sin^2\theta}{V^2(\theta)}\Bigr\rangle_\theta},
\end{equation}
which is the same formula as $\omega_{\rm m}$. On the other hand, $\bigl\langle J_{Q_{\rm b}}^{(i)}\bigr\rangle$ can be written as
\begin{align}
\bigl\langle J_{Q_{\rm b}}^{(i)}\bigr\rangle&=\lim_{\tau\to\infty}\frac{1}{\tau}\int_{0}^{\tau}G\frac{1+\sin\theta_i}{2}(T_{\rm b}-T(\theta_i,\omega_i))dt\\
&=
\lim_{\tau\to\infty}\frac{1}{\tau}\int_{0}^{\tau}G\frac{1+\sin\theta_i}{2}(T_{\rm b}-T_{{\rm eff}}(\theta_i))dt
+\lim_{\tau\to\infty}\frac{1}{\tau}\int_{0}^{\tau}\frac{1+\sin\theta_i}{2}\frac{\sigma\sin\theta_i}{V(\theta)}\omega_idt+\bigl\langle\mathcal{O}(\Delta T\omega_i,\omega_i^2)\bigr\rangle\label{heat flux}.
\end{align}
The first and second terms of Eq. (\ref{heat flux}) are calculated as follows:
\begin{align}
&\lim_{\tau\to\infty}\frac{1}{\tau}\int_{0}^{\tau}G\frac{1+\sin\theta_i}{2}(T_{\rm b}-T_{{\rm eff}}(\theta_i))dt\\
&=\frac{G\Delta T}{4}\lim_{\tau\to\infty}\frac{1}{\tau}\int_0^{\tau}\cos^2\theta_idt\\
&=\frac{G\Delta T}{4}\lim_{N\to\infty}\frac{1}{\langle\omega_i\rangle\sum_{k=1}^N\tau_k^{(i)}}\int_0^{\sum_{k=1}^N\tau_k^{(i)}}\cos^2\theta_i\langle\omega_i\rangle dt\\
&\approx
\frac{G\Delta T}{4}\lim_{N\to\infty}\frac{1}{2\pi N}\int_0^{2\pi N}\cos^2\theta d\theta+\bigl\langle\mathcal{O}(\Delta\theta_i)\bigr\rangle\\
&=\frac{G}{8}\Delta T+\bigl\langle\mathcal{O}(\Delta\theta_i)\bigr\rangle\label{JO},
\end{align}
\begin{align}
&\lim_{\tau\to\infty}\frac{1}{\tau}\int_{0}^{\tau}\frac{1+\sin\theta_i}{2}\frac{\sigma\sin\theta_i}{V(\theta)}\omega_i dt\\
&=
\frac{\sigma}{2}\left(\lim_{N\to\infty}\frac{1}{\sum_{k=1}^{N}\tau_k^{(i)}}\int_0^{2\pi N}\frac{\left(1+\sin\theta_i\right)\sin\theta_i}{V(\theta_i)}d\theta\right)\\
&=\frac{\sigma}{2}\left(\frac{1}{2\pi}\int_0^{2\pi}\frac{(1+\sin\theta)\sin\theta}{V(\theta)}d\theta\right)\left(\lim_{N\to\infty}\frac{2\pi N}{\sum_{k=1}^N\tau_k^{(i)}}\right)\\
&=\frac{\sigma}{2}\Biggl\langle\frac{\sin^2\theta}{V(\theta)}\Biggr\rangle_\theta\langle\omega_i\rangle.
\end{align}
It is then straightforward to obtain the quasi-linear relation w.r.t. $\bigl\langle J_{Q_{\rm b}}^{(1)}\bigr\rangle+\bigl\langle J_{Q_{\rm b}}^{(2)}\bigr\rangle$ by neglecting higher order terms in Eqs. (\ref{heat flux}) and (\ref{JO}).

\bibliographystyle{junsrt}
\bibliography{Yin_20231016}

\newpage

\end{document}